\documentclass[reprint,fleqn,superscriptaddress,twocolumn,showpacs,
amsmath,amssymb,prb,aps,short bibliography]{revtex4-1}
\usepackage[all]{xy}
\usepackage{gensymb}
\usepackage{graphicx,psfrag,times,epsfig,color}
\usepackage{verbatim,natbib}
\usepackage{float}
\usepackage{color}
\usepackage{makeidx}
\usepackage{amsmath}
\usepackage{bm}
\usepackage{amsfonts}
\usepackage{amssymb}
\usepackage{hyperref}

\begin{document}

\title{Calculations of quantum oscillations in cuprate superconductors
 considering the pseudogap}
\author{E. V. L. de Mello}
\affiliation{Instituto de F\'{\i}sica, Universidade Federal Fluminense, 24210-346 Niter\'oi, RJ, Brazil}

\email[Corresponding author: ]{evandro@mail.if.uff.br}

\begin{abstract}
The observations of quantum oscillations frequencies in overdoped cuprates were in
agreement with a charge density contained in a cylindrical Fermi 
surface but the measured frequencies of underdoped compounds were much smaller than expected. 
This was attributed to a topological transition into small pockets of Fermi surface 
associated with the existence of charge density waves. 
On the other hand, 
spectroscopic measurements suggested that the large two-dimensional Fermi surface changes
continuously into a set of four disconnected arcs. 
Here we take into account the effect of the pseudogap
that limits the available $k$-space area where the Landau levels are developed
on the Luttinger theorem and obtain the correct total carrier densities. 
The calculations show how 
the disconnected arcs evolve into a closed Fermi surface reconciling  
the experiments.

\end{abstract}

\maketitle


The details of different Fermi surfaces in metals have long 
been explored by  quantum oscillations (QO) experiments, like conventional 
de Haas-van Alphen effect. The 
quantum oscillations are a direct consequence of the quantization of closed Landau orbits 
perpendicular to an applied  magnetic
field. The energy of a given orbit depends on the applied field, and varying the field, 
there is always a resonance when a Landau level crosses the Fermi energy,
from which the transverse (to the field) area of the Fermi surface is derived. The QO
in the overdoped cuprate superconductor Tl$_2$Ba$_2$CuO$_{2+\delta}$ (Tl2201) 
show the existence  of a large Fermi surface covering nearly two-thirds of the Brillouin zone (BZ)\cite{Vignolle2008} 
in very good agreement with the angle-resolved photoemission spectroscopy
(ARPES) results\cite{Plate2005}.  The frequency ($F$) of
oscillation is related with cross-sectional area $A_{\bf{k}}$ of the Fermi surface by the Onsager relation,
$A_{\bf{k}} = 2\pi e F/\hbar$. Assuming that the Fermi surface is strictly two-dimensional, then the total itinerant carrier
density per plane per area is given by Luttinger's theorem; 
$n = 2A_{\bf{k}}/A_{\rm BZ} = 1+p$, where
$A_{\rm BZ} = (2\pi/a)^2 \approx 265$ nm$^{-2}$ is the BZ area, $p$ is the average doping and
$a$ the lattice parameter. These relations work quite
well with several overdoped Tl2201 compounds\cite{Vignolle2008,Plate2005,Bangura2010} 
with $p \approx 0.25-0.30$.

However QO measurements in underdoped YBa$_2$Cu$_3$O$_y$ (Y123) and in 
HgBa$_2$CuO$_{4+y}$ (Hg1201) are both 25-32 times 
lower than the overdoped Tl2201 frequency\cite{NDL2007,Bangura2008,Yelland2008,Jaudet2008,Barisic2013,Chan2016},
implying, by the Onsager
relation, in very small cross-sectional areas $A_{\bf{k}}$. Consequently, 
there is a discrepancy with 
Luttinger theorem since typical doping differences are lower only by a factor of 2-3. 
To explain this overdoped/underdoped difference it was proposed
a Fermi surface reconstruction into several small pockets of Fermi 
surface\cite{NDL2007,Bangura2008,Yelland2008,Jaudet2008,Barisic2013,Chan2016}.
The change of sign in the Hall resistances with the 
temperature in high magnetic-field-induced normal
state of (Y123) and on Hg1201 suggested that these pockets are electron-like rather than hole-like\cite{LeBoeuf2007,Doiron-Leyraud2013}. 
Since the negative Hall resistances occur between $p =0.07-0.15$\cite{Badoux2016},
their existence and the crossover between electron-hole pockets were attributed 
to the incommensurate charge order (CO) phase\cite{LeBoeuf2007,Taillefer2009}
or charge density waves (CDW) superlattice formation\cite{Chakravarty2008}. 
Theoretical approaches beyond semiclassical approximations taking the effects of CO 
into consideration obtained signatures of electron pockets\cite{Allais2014} in
qualitative agreement with the experimental data.
 
In order to check this Fermi surface reconstruction crossover 
an effort was made to perform angle resolved photoemission spectroscopy
(ARPES) experiments with Y123\cite{Hossain2008} and Hg1201\cite{Vishik2014}, 
but they both did not detected the presence of electron pockets. Thus, the 
interpretation of QO in terms of electron and hole pockets differs markedly from
single-particle spectroscopy, suggesting that high magnetic fields might
induce a new electronic state. 

We provide a new interpretation to QO experiments taking into account the
charge instabilities on Luttinger theorem, that was
originally derived in the context of a Fermi liquid with uniform density.
In previous papers\cite{deMello2012,Mello2017,Mello2019a} we discussed the large  
amount of experimental evidences for spontaneous symmetry breaking and anomalous 
long-range ordered electronic states arising near the pseudogap (PG) temperature 
$T^*(p)$. The correlations between charge modulations
wavelengths $Q_{\rm CO}$ ($= 1/\lambda_{\rm CO}$) in real space 
and the distance between the Fermi arcs tips in $k$-space that is
dominated by the PG was established by scanning tunneling microscopy 
(STM)\cite{Shen2005,Wise2008} and by a combination
of ARPES, STM and resonant x-rays (REXS)
on Bi$_2$Sr$_{2-{\rm x}}$La$_{\rm x}$O$_{6+\delta}$ (Bi2201)\cite{Comin2014}. 
On the theoretical side, we demonstrated that the PG energy
$\Delta_{\rm PG}(p)$ is proportional to the ground state energy of a two-dimensional
potential well with radius equal to $\lambda_{\rm CO}(p)$\cite{Mello2019a}, 
establishing a direct connection between the CO spatial scale and the PG energy 
$\Delta_{\rm PG}(p)$.
The charge anomalies are strong evidence of a non-Fermi liquid behavior 
and pose the question of how does it affect the Luttinger theorem? We shall 
answer this question in the next paragraphs.

\begin{figure}
\includegraphics[height=4.0cm]{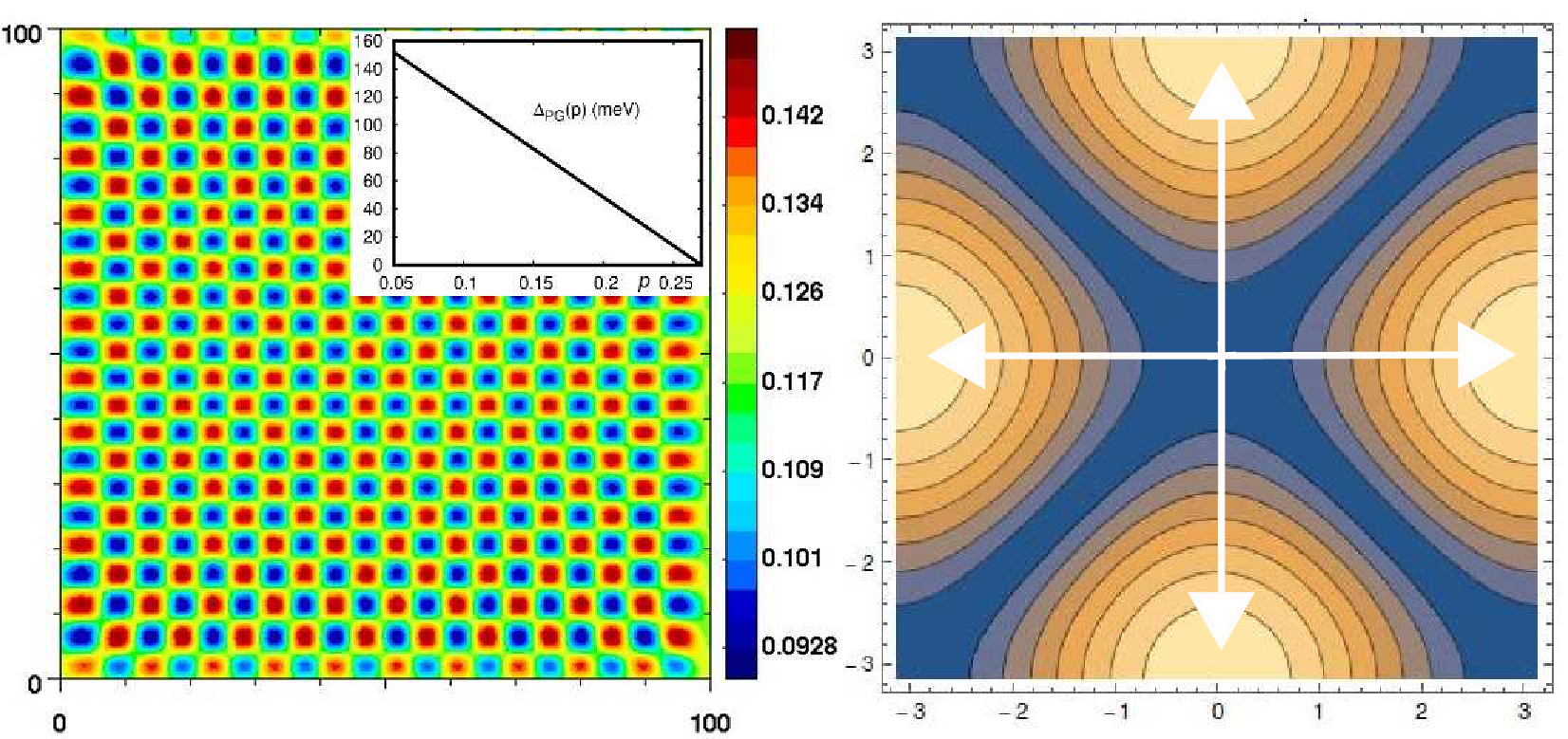}
\caption{ The effect of the PG in real and $k$-space.
(a) Checkerboard charge density map simulation for $p = 0.12$ Bi2212 compound
with hole-rich domains in red and hole-poor in blue. 
The inset shows the experimental linearly decreasing $ \Delta_{\rm PG}(p, 0)$ energy
amplitude in meV as function of $p$\cite{Huefner2007,Raman2011,DamascelliRMP2003}. In
(b) we show the constant energy $d$-wave pseudogap $|\Delta_{\rm PG}(k_x,k_y)|$
color cuts which vanishes at the BZ diagonals $k_x = \pm k_y$, increases along
the white arrows or nodal directions reaching the maxima at the antinodes.
The dark blue region along the diagonals exemplifies a
RBZ area (Eq. \ref{RBZ}) since it is similar to the area inside the $p = 0.10$ curves labeled ``1'' in Fig. (\ref{figLL})(a) and (b).
}
\label{fig1PG}
\end{figure}

 Another important property comes from measurements 
of the momentum or wave vector direction
like Raman scattering\cite{Huefner2007,Raman2011} and ARPES\cite{DamascelliRMP2003}
that established the $d$-wave symmetry of the $\Delta_{\rm PG}(p)$. Because of their connection, 
we simulate the CO anomalies by a Cahn-Hilliard (CH) approach in real space 
that is quite precise in reproducing the experimental observations
\cite{deMello2009,deMelloKasal2012,deMello2012,Mello2017}.
In  Fig. \ref{fig1PG}(a) we show an example of simulation; the $p = 0.12$ compound
from the Bi2212 family exhibiting CO with planar checkerboard structure in 100 vs. 100 
unit cells similar to what we have done previously\cite{Mello2017}. 
Fig. \ref{fig1PG}(b) is a plot made by Mathematica that displays
constant energy cuts of the two-dimensional $k$-space single particle energy 
$\Delta_{\rm PG}(p, T)(k_x,k_y) = \Delta_{\rm PG}(p, T)|cos(k_xa)-cos(k_ya)|/2$. 
It vanishes near the (dark blue) diagonals or nodal directions 
and increases in the antinodal directions indicated by the white arrows pointing to larger 
energies represented by lighter colors.
A simple two-dimensional integral done also
with Mathematica yields the area inside any these constant energy cuts.
On the other hand, $\Delta_{\rm PG}(p, 0)$ decreases linearly with doping\cite{Huefner2007,Raman2011,DamascelliRMP2003} as shown in the inset of 
Fig. \ref{fig1PG}(a) what implies generally in smaller $\Delta_{\rm PG}(p, T)$, that 
is, the constant energies curves are displace toward the antinodes Fig. \ref{fig1PG}(b) 
for overdoped materials.
Consequently, the charges in underdoped compounds have larger energy
constraints and their motion in $k$-space are limited to the nodal directions. 
In summary, the charge modulations
in real space and energy restriction in momentum space are two manifestations of 
the PG effect in cuprates and we show below how to take into consideration 
in the Luttinger theorem.
We have previously developed a theory that, at
low temperatures, the two components SC order parameter (amplitude 
$\Delta_{\rm sc}$ and phase $\Phi$) 
is induced by lattice fluctuations in alternating
CDW or CO domains\cite{deMello2014,Mello2017}.
The local SC phases are coupled by Josephson energy 
$E_{\rm J}^{\rm ij}(T) = (\Delta_{\rm sc}(r_{\rm i},T), \Delta_{\rm sc}(r_{\rm j},T), R_{\rm n}),T)$, 
where $\Delta_{\rm sc}(r_{\rm i},T)$
is the SC amplitude in the charge domain $i$ and $R_{\rm n}$ is the total resistance
just above the SC transition
like in granular\cite{Mello2017,Mello2019a} or disordered superconductors\cite{Spivak1991}.
The $E_{\rm J}^{\rm ij}(T)$ are the local superfluid
phase stiffness\cite{Spivak1991} $\rho_{\rm sc}^{\rm ij}$ and its average
over the whole plane ${\left <E_{\rm J} \right >} = \rho_{\rm sc}$ is 
directly proportional to the planar or two-dimensional
superfluid density. In general, $n_{\rm sc} = (4 k_{\rm B} m^* / \hbar^2 ) \rho_{\rm sc}$
where $m^*$ is the effective mass of the electron\cite{Bozovic2016}.

Now, under the extreme QO conditions; applied fields around 50 T and temperature 
$T \approx 1$ K, the local Cooper pairs or the SC amplitudes break down and the 
superfluid density $n_{\rm sc}$ become low energy normal 
carriers,  that is, $n_{\rm sc} = n $. As mentioned above,
the superfluid phase stiffness $\rho_{\rm sc}$ gives this charge density and
more, it determines also the energy scale or available kinetic energy
of these unpaired holes. On the other hand,
the PG or CDW charge instabilities are unaffected by a
50 T magnetic fields as demonstrated, for instance, by Chang{\it et al}\cite{Chang2012}.
Consequently, in QO experiments the free carriers from unpaired holes 
$\approx n$ are constrained by the PG
to obey the following relation in the BZ,
\begin{equation}
\rho_{\rm sc}(p, T=0$ K$) \ge |\Delta_{\rm PG}(p, 0)(k_x,k_y)| .
\label{RBZ}
\end{equation}
This inequality defines the two-dimensional {\it restricted BZ area} (RBZ($p$)) for 
particles under QO conditions and charge instabilities or in the  
presence of the PG, a restriction that does not exist for free particles.
It means that the particles originally from the SC condensate 
are restricted to the region where the PG
is weaker and, therefore it may not be energetically favorable to be near the 
anti-nodes $(k_x,k_y) =(\pm \pi/a, 0)$ or $(0, \pm \pi/a)$ regions.
To illustrate, the RBZ for underdoped compounds that $|\Delta_{\rm PG}|$ is generally
much larger than $\rho_{\rm sc}$ resembles the blue region near 
the BZ diagonals in Fig. \ref{fig1PG}(b).
As $|\Delta_{\rm PG}|$ becomes small with increasing doping, the RBZ area
becomes larger, eventually comprising the dark blue, the light blue, the dark brown and
the lighter brown regions in Fig. \ref{fig1PG}(b) and when $|\Delta_{\rm PG}| \rightarrow 0$ 
the RBZ becomes the full $A_{\rm BZ} = (2\pi/a)^2 \approx 265$ nm$^{-2}$ BZ area.

\begin{figure}[H]
\includegraphics[height=4.0cm]{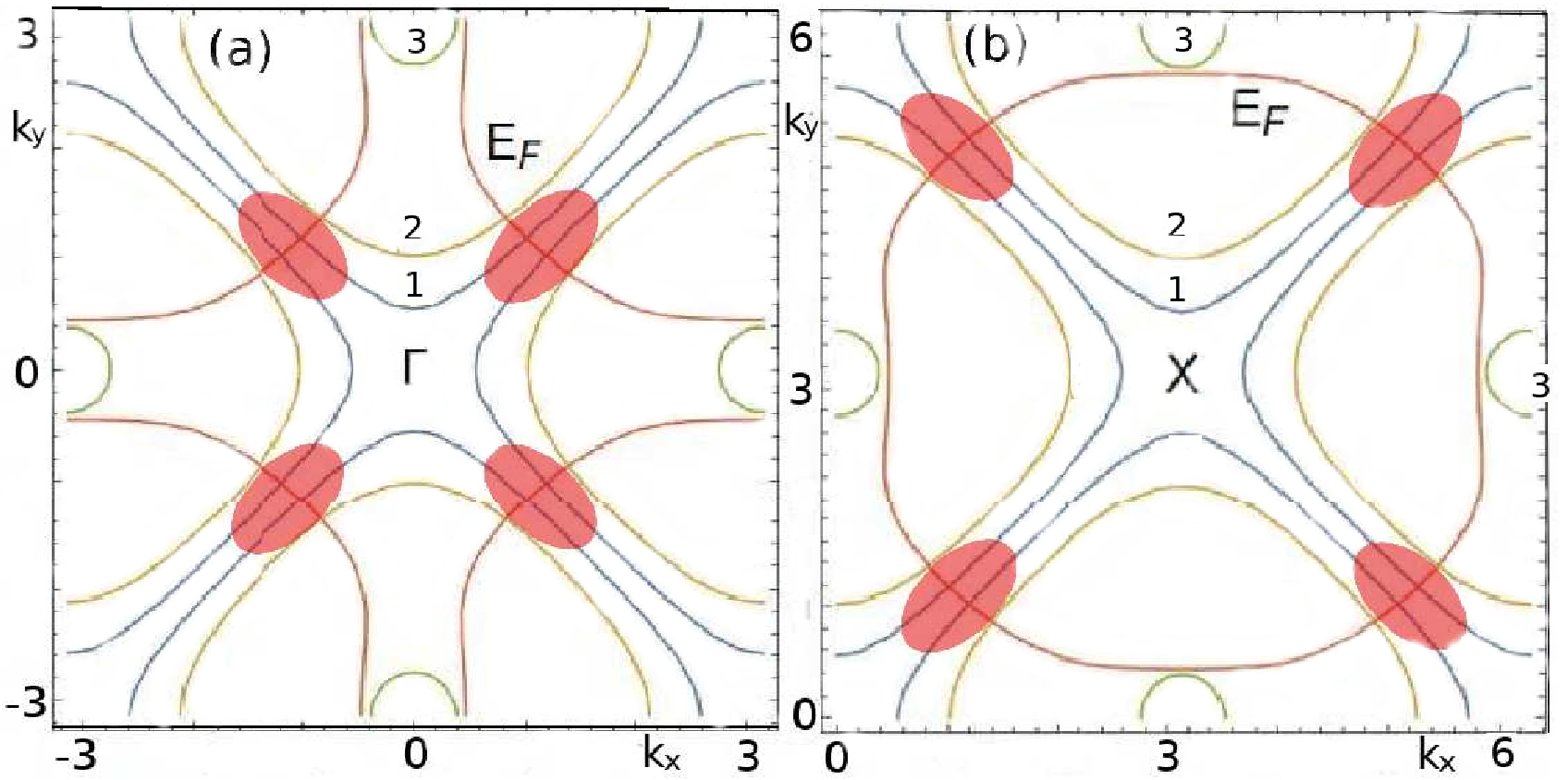}
\caption
{ Three curves $\rho_{\rm sc}(p, 0) = |\Delta_{\rm PG}(p, 0)(k_x,k_y)|$,
for  two Y123 underdoped compounds with $p = 0.10$, labeled ``1'', 0.125 ``2'' and 
a Tl2201 overdoped $p = 0.27$ labeled ``3'' in the BZ for electrons (a) 
and in (b) for holes.
The area inside the four branches of each curve, that is, from the lines toward the 
center $\Gamma$ in (a) or X in (b) defines the RBZ.
We draw also the $E_{\rm F}$ for $p = 0.27$ for comparison that is nearly
the same for the other dopings. Under the QO conditions the free particles from
underdoped superconductors cannot move around the $E_{\rm F}$ because 
the four PG branches act like strong
potential barriers confining the carriers near the zone diagonals.
We represent these constrained Landau levels schematically
by the squeezed red circles or ellipsoids whose areas obey the Onsager relation.
}
\label{figLL}
\end{figure}

To calculate explicitly the RBZ from Eq. (\ref{RBZ}), we obtain the $\rho_{\rm sc}(T)$ 
values for Y123 and other cuprates from our previous paper\cite{Mello2019b} and
$ \Delta_{\rm PG}(p, 0)$ are taken from the experiments\cite{Huefner2007,DamascelliRMP2003}.
They are listed in Table I. 
We use Mathematica to plot the two-dimensional curves 
$\rho_{\rm sc}(p, 0) = |\Delta_{\rm PG}(p, 0)(k_x,k_y)|$ for two 
Y123 underdoped compounds with 
$p = 0.10 $ labeled by ``1'', 0.125 by ``2'' and one Tl2201 overdoped with $p = 0.27$ 
by ``3'' in Fig. \ref{figLL}(a) for electrons and Fig. \ref{figLL}(b) for holes.
We use also Mathematica to integrate directly the two-dimensional areas 
below these three curves the defines the RBZ. The areas or RBZ in the curves  ``1'' 
and ``2'' in unit  of nm$^{-2}$ are listed in Table I.
In the far overdoped region $\Delta_{\rm PG}(p,0) \rightarrow 0$ and, in this case,
the particles from the SC condensate are not restricted to the diagonals and
can occupy the entire BZ. 
Therefore the Landau
levels at the Fermi surface resembles the free particle case and 
go around describing almost perfect circles, like
Tl2201 with $p = 0.27$ shown by the red curve labeled 
by ``3'' in Fig. \ref{figLL} (b), in total agreement 
with the experimental results\cite{Vignolle2008}. For $p < 0.27$, 
$\Delta_{\rm PG}(p, 0)$ increases as $p$ decreases and according
the above discussion, the particles at the Fermi energy cannot
penetrate in the antinodal region and are forced to stay near the BZ diagonals.
In other words, the holes  are confined to the nodal 
or diagonal directions as it is the case for $p = 0.10$ and 0.125. Under QO
conditions of very high magnetic 
fields the particles can move only in small circles or ellipsoids
around the four nodal portions or arcs of the Fermi surface
as schematically displayed in Fig. \ref{figLL} (a) and (b).
We emphasize that we do not calculate the areas of these circles or
ellipsoids, but their areas $A_k$ are obtained from the QO 
measured frequencies $F(p)$\cite{NDL2007,Bangura2008,Yelland2008,Jaudet2008,Barisic2013,Chan2016}
and the Onsager's relation and listed in Table I.

\begin{table*}[t!]
\caption{
We list the data and calculations of the four compounds used in QO experiments\cite{NDL2007,Bangura2008,Yelland2008,Jaudet2008,Barisic2013,Chan2016}.
The measured frequencies in the third column are in tetrahertz (T) and
the respective areas from the Onsager relation are in the fourth column. $\rho_{\rm sf}(0)$ for Y123 is from the previous paper\cite{Mello2019b}, Tl1201 is from 
experiments\cite{LscoOver1993} and the value of Hg1201
is assumed to be close to that of Bi2212\cite{Mello2019b}. The
RBZ areas inside curves labeled by 1 and 2 of Fig. \ref{figLL}, 
defined by Eq. \ref{RBZ}, are calculated directly simply integrating
with Mathematica the area inside these two plots. Notice that for $p = 0.30$ in the 
last line the PG is zero the RBZ is equal the BZ area $ = 265 $ nm$^{-2}$.
The theoretical
densities from Eq. \ref{ModLutt} are in the last
column and in agreement with the experimental values of column 1.
}
\label{table1}
 \begin{tabular}{|c|c|c|c|c|c|c|c|}\hline \hline
$p$ & $T_{\rm c}$ & $F(p)$ (T) &  $A_{\bf{k}}$ (nm$^{-2}$) & $\Delta_{\rm PG}$ (meV) & $\rho_{\rm sc}(0)$ (K)& RBZ nm$^{-2}$ &  $1+p$ (Eq. \ref{ModLutt})    \\ \hline

0.10 (Y123)& 57 K & $530 \pm 20$ T [\onlinecite{NDL2007}] & 5.10 &  100.0 meV [\onlinecite{Huefner2007}]& 87.5 & 36.89 (1) & 1.106\\ \hline

0.125 (Y123)& 64 K & $660 \pm 15$ T [\onlinecite{Bangura2008,Yelland2008}] & 6.37 & 91.0 meV[\onlinecite{Huefner2007}] & 92.5 & 45.26 (2) & 1.126\\ \hline

0.090 (Hg1201)& 72 K & $840 \pm 30$ T [\onlinecite{Barisic2013}]& 8.08 & 100.0 meV [\onlinecite{RamanHg2013}]& 143.4 & 59.2 &  1.091\\ \hline

0.30 (Tl2201)& 10 K & $18100 \pm 50$ T [\onlinecite{Vignolle2008}]& 172.8 & 4 meV [\onlinecite{Huefner2007}] &29.5  & 265 (3) & 1.303\\ \hline
 \end{tabular}
 \end{table*}

\begin{figure}
\includegraphics[height=4.0cm]{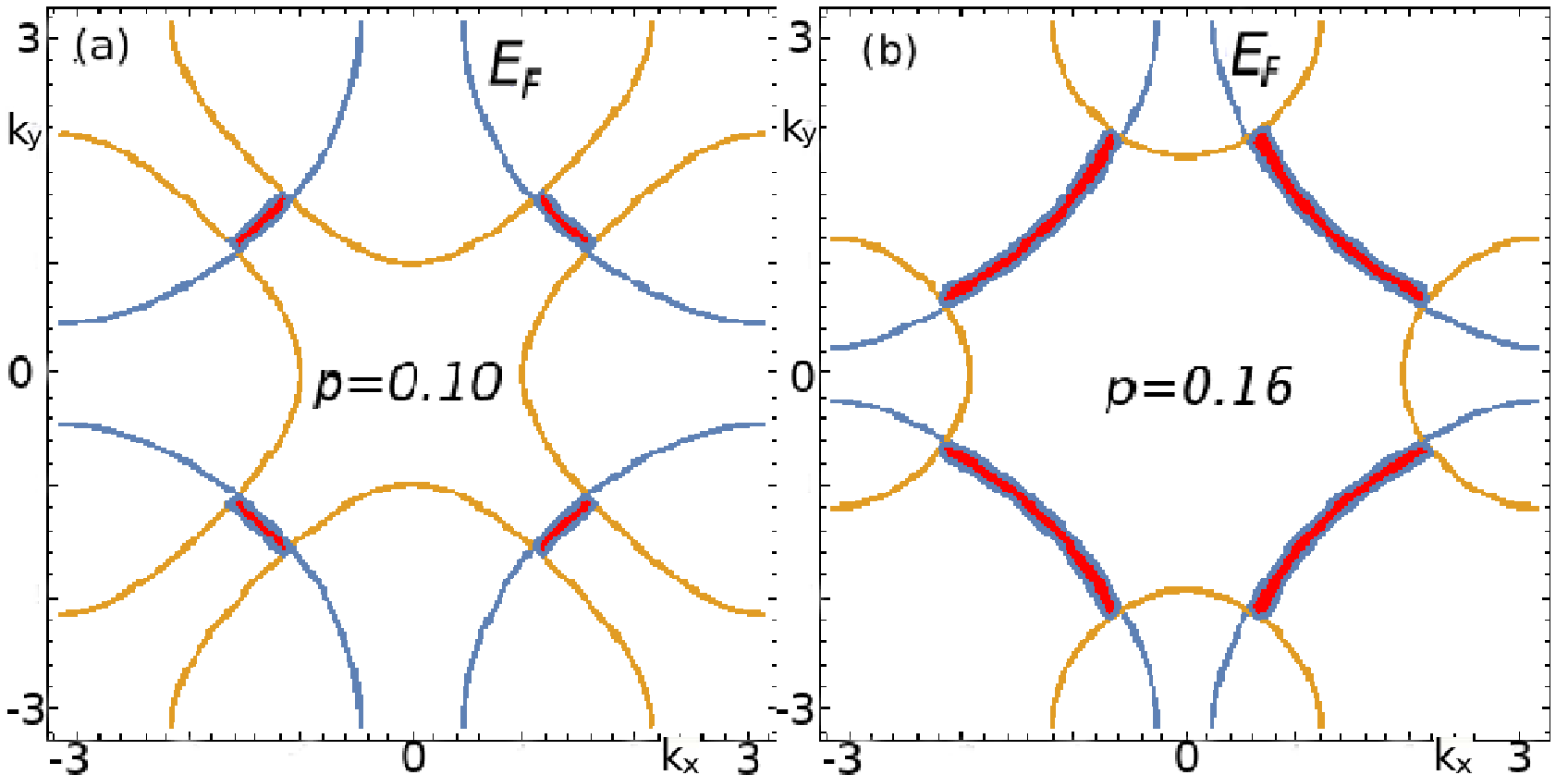}
\caption{ 
The $d$-wave SC amplitude $|\Delta_{\rm sc}(k_x,k_y)|$ at $T \approx T_{\rm c} + 10$ K.
The same functional form of the PG and SC gap is an additional evidence of the connection
between them. At these temperatures the size of the Fermi arcs or 
gapless regions are determined 
by thermal and quantum phase fluctuations, they increase and cover the 
whole Fermi surface when $T \rightarrow T^*$. 
To calculate the gapless arc sizes it is needed to take these two fluctuation effects
into consideration like the derivations for three Bi2212 compounds in
Fig. 6 of Ref. \onlinecite{Mello2017}, after ARPES measurements\cite{Lee2007}.
}
\label{Farcs}
\end{figure}
Consequently, the measured QO frequencies $F$ on underdoped samples come from the 
holes oscillations around these four separated regions or Fermi arcs bound by
$|\Delta_{\rm PG}(p, 0)(k_x,k_y)|$ shown in Fig. \ref{figLL}(a) and (b).
The four contributions are alike and only one single frequency is measured experimentally
yielding the area $A_{\bf{k}}(p)$ from the Onsager relation. 
Taking all the above into consideration, we rewrite Luttinger theorem as: 
\begin{equation}
n = 1 + p = A \times 2A_{\bf{k}}(p)/$RBZ$(p) , 
\label{ModLutt}
\end{equation}
where A = 1 for $p \ge 0.27$ and 4 otherwise to account for the four PG nodal
regions that cross the Fermi level. As we have already explained, we calculate 
RBZ$(p)$ with the 
zero temperature $\Delta_{\rm PG}(p, 0)$\cite{Huefner2007} 
and $\rho_{sc}(p, 0)$ from our previous paper\cite{Mello2019b} and from
experiments\cite{LscoOver1993}.
The $\Delta_{\rm PG}(p, 0)$ and $\rho_{\rm sc}(p, 0)$ 
values used in the calculations
for $p = 0.09$ (Hg1201), 0.010, 0.125 (Y123) and $p = 0.30$ (Tl2201)
are all listed in Table \ref{table1}. We do not have data to calculate
$\rho_{\rm sc}(0.09)$ for Hg1201 and we used that
of Bi2212 in the preceding paper\cite{Mello2019b} because their
similar $T_{\rm c}(p)$ curves. 
It is known that $\rho_{\rm sc}(p, 0)$
is maximum near or further the optimum doping and is small in the underdoped region
while $\Delta_{\rm PG}(p, 0)$ decreases linearly with $p$. Both effects 
combined imply that RBZ$(p)$ decreases rapidly as $p \rightarrow 0$ and 
are much less than the full BZ area also listed in Table I.
Using the areas $A_{\bf{k}}(p)$ derived from the QO measured frequencies $F$ 
by the Onsager relation and Eq. (\ref{ModLutt}), we 
derive hole doping densities in very good
agreement with the compounds used in the QO experiments
\cite{Vignolle2008,Plate2005, Bangura2010,NDL2007,Bangura2008,Yelland2008,Jaudet2008,Barisic2013,Chan2016} as listed in Table I. 

It is  important to emphasize that the Landau levels
at the Fermi surface constrained by the PG provides also an explanation to the existence 
of the four-hole pockets with the measured frequencies or areas.
Our approach is also in qualitative agreement with the Fermi
arcs measured by several ARPES experiments in the absence of a magnetic field\cite{Norman1998,Plate2005,Shen2005,Lee2007,Yoshida2006,Kanigel2006,Yoshida2012}.
We have previously demonstrated\cite{Mello2017} that the 
Fermi arcs become finite at $ T > T_{\rm c}$ due to thermal and quantum
fluctuations of the local superconducting order parameter
phase $\Phi( r_i)$. We have shown that the average superconducting amplitude 
${\left <\Delta_{\rm sc} \right >}$ is finite but
the order parameter vanishes along the nodal direction due to average quantum and 
thermal phase fluctuations\cite{Mello2017}.
In Fig. \ref{Farcs} we show the results of these phase fluctuation 
calculations for $p = 0.10$ (a) and 0.16 (b) at $T \approx T_{\rm c} + 10$ K 
that reproduced ARPES measurements 
on Bi2212\cite{Lee2007}. We should mention that previous works connecting
QO and CO anomalies that also reconciled QO and ARPES obtained 
only qualitative agreement with the experimental trends on underdoped compounds\cite{Allais2014}
while we present detailed quantitative results for any doping level.

In summary, we demonstrated that taking the role of the PG in QO
experiments we can understand why the measured frequencies $F$ that are generally
proportional to the charge densities change by factor ten times larger than 
the average doping levels.
We showed how low energy carriers, reminiscent from broken Cooper
pairs under strong magnetic fields (the QO conditions) are 
constrained by the PG (or CDW instabilities)
to the regions near the diagonals of the BZ. This new topological or geometric
concept of a restricted BZ (RBZ) is the central idea of our work.
This restriction in the BZ area modifies
completely Luttinger's theorem that assumes the whole two dimensional 
BZ as the domain to the electrons on the CuO planes. The 
$d$-wave PG $|\Delta_{\rm PG}(k_x,k_y)|$ plays the role of a two- dimensional
non-constant potential in the BZ and the superfluid phase stiffness 
$\rho_{\rm sf}(p, 0)$ yields the energy scale to the carriers. We can think
that the PG is the height of four symmetric mountains around a lake
and $\rho_{\rm sf}$ is the level of the lake. 
We stress that this correction to the Luttinger's theorem is
our main contributions to the understand the QO results and 
we do not calculate any frequency as function of
magnetic field although strong fields are essential to
transform the superfluid density in free particles.
The RBZ  in connection with the areas derived from
the QO measured frequencies $F(p)$ through the Onsager's relations yields the correct
hole doping $p$. 
The calculations provided clear quantitative results to the hole densities 
(listed in Table I), a 
new interpretation to the QO experiments and indicate a way to reconcile these 
observations with the Fermi arcs measurements by ARPES.

I acknowledge partial support by the Brazilian agencies CNPq and
FAPERJ.

\end{document}